\documentclass[12pt]{article}%

\usepackage{amssymb}
\usepackage{amsmath}
\usepackage{graphicx}%
\usepackage{amsfonts}
\usepackage{authblk}

\usepackage{bm}
\usepackage{epsfig}
\usepackage{color}
\usepackage[body={17.5cm,24cm},top=1.75cm]{geometry}
\usepackage[linktoc=all]{hyperref}
\usepackage{ulem,color}
\usepackage{soul}
\usepackage[toc,page]{appendix}
\usepackage[usenames,dvipsnames,svgnames,table]{xcolor}

\begin{document}

\title{\Large \bf Impact of the $^7$Be($\alpha, \gamma$)$^{11}$C reaction on the primordial abundance of $^7$Li}

%\correspondingauthor{Shubhchintak}
%\email{shub.shubhchintak@tamuc.edu}

\author{M. Hartos}
\affil{Department of Physics and Astronomy, Texas A$\&$M University-Commerce, Commerce, TX 75429, USA}

\author{C. A.  Bertulani}
\affil{Department of Physics and Astronomy, Texas A$\&$M University-Commerce, Commerce, TX 75429, USA}

\author{Shubhchintak}
\affil{Department of Physics and Astronomy, Texas A$\&$M University-Commerce, Commerce, TX 75429, USA}

\author{A. M. Mukhamedzhanov}
\affil{Cyclotron Institute, Texas A$\&$M University, College Station, TX 77843, USA}

\author{S. Hou}
\affil{Institute of Modern Physics, Chinese Academy of Sciences, Lanzhou, 730000, China}

\maketitle

\begin{abstract}

We calculate the radiative capture cross section for $^7$Be($\alpha, \gamma$)$^{11}$C and its reaction rate of relevance for the big bang nucleosynthesis. The impact of this reaction on the primordial $^7$Li abundance is revised including narrow and broad resonances in the pertinent energy region. Our calculations show that  it is unlikely  that very low energy resonances in $^{11}$C of relevance for the big bang nucleosynthesis would emerge within a two-body potential model. Based on our results and a comparison with previous theoretical and experimental analyses, we conclude that the impact of this reaction on the so-called ``cosmological lithium puzzle" is completely irrelevant. 

\end{abstract}


\section{Introduction} \label{sec:intro}

Since it was devised more than half a century ago, the successes of the big bang theory turned it into a powerful tool to probe the physics in the early universe and allowed one to explore phenomena beyond the standard model of particle physics. The big bang model assumes an isotropic and homogeneous universe with a radiation dominated expansion. The only parameter of the theory is the baryon number density to photon number density ratio $\eta_B$, which can be determined with a high precision by the analysis of the anisotropies of the cosmic microwave background (CMB). Such anisotropies have been observed with the Wilkinson Microwave Anisotropy Probe (WMAP) \cite{komatsu} and with the Planck mission \cite{planck}. As a plethora of data were disclosed, including nuclear abundances and other astronomical observations, it was found that the big bang model predictions agree with observations of light nuclei abundances such as deuterium and helium. However, the Big Bang Nucleosynthesis (BBN) prediction for the $^7$Li abundance is about 3 times larger than the observations \cite{cyburt, coc04}. In fact, the lithium abundance [$^7$Li/H]$_{\text{BBN}}$ is predicted to be $5.12^{+0.71}_{-0.62}\times10^{-10}$ \cite{cyburt} if one uses $\eta_B$ deduced by WMAP observations \cite{komatsu} and  $4.89^{+0.41}_{-0.39}\times10^{-10}$ \cite{coc04} if one uses $\eta_B$ derived from Planck results \cite{planck}. However,  the most recently observed value for [$^7$Li/H] in metal-poor halo stars is $1.58^{+0.35}_{-0.28}\times10^{-10}$ \cite{sbordone}.  
This discrepancy is known in the literature as the cosmological lithium problem. For about two decades, various theoretical efforts to solve this problem have been undertaken but so far no acceptable justification was found within nuclear physics. Non-nuclear physics solutions, including models beyond the standard BBN have been reported \cite{pospelov,coc13,kusakabe,yamazaki,goudelis,suqing1,daniel}, although there are so far no experimental evidences to support these models.

After nucleosynthesis during the big bang occurred, $^7$Li was synthesized by other processes during novae, pulsations of asymptotic giant branch stars and by spallation reactions of cosmic-rays with carbon and oxygen \cite{reeves}. Low metallicity stars are thus considered ideal sites for detecting primordial Li abundance and, to date, almost all the observations focus on metal-poor stars in the halo of our galaxy. The value of [$^7$Li/H] from the evaluation of Sbordone et al. \cite{sbordone} is adopted extensively as the standard value of the  primordial lithium abundance. BBN calculations with a nuclear physics focus have been reported in several works which can be divided into two categories; nuclear reactions to create $^7$Li and those destroying $^7$Li \cite{Lam17}. BBN predicts that the the majority of the primordial $^7$Li production arises from the $^7$Be decay by electron capture during the two months after BBN stops. Thus, for the solution of the Li problem,  reactions involving $^7$Be could be more significant than those involving $^7$Li.

In Ref. \cite{coc10}, it was mentioned that the $^3$He($\alpha$,$\gamma$)$^7$Be and $^7$Be(n, p)$^7$Li reactions  are the leading reactions for the production and destruction of $^7$Be. Many studies have focused on these two reactions during the past few years and will probably continue in the near future. However,  recent theoretical and experimental investigations suggest that these reactions may not solve the lithium problem \cite{boer,coc17,broggini12}.   $^7$Be($n, \alpha$)$^4$He is another neutron capture reaction responsible for the destruction of $^7$Be which has also been investigated in several works \cite{hou, barba,kawa, lamia}. But the results show that it could only worsen the lithium problem. 

Despite of several failed attempts, studies based on the nuclear physics input for the BBN are still ongoing. Ref. \cite{broggini12}   lists a series of nuclear reactions and analyze their role in solving the lithium problem. Almost all of the potential solutions considered in that work have been ruled out by their theoretical analysis, except maybe for the $^7$Be($\alpha,\gamma$)$^{11}$C reaction. The authors speculate that an yet to be observed narrow resonance in the low energy region in $^{11}$C could significantly reduce the $^7$Li BBN abundance \cite{broggini12}. Similar speculations have been  also presented in Ref. \cite{civit} but with a relatively wide resonance. On the other hand, an experiment reported in Ref. \cite{hammache} probably suggests that no resonance exists in the low energy region of this system. 

Theoretical calculations for the cross section of $^7$Be($\alpha, \gamma$)$^{11}$C reaction at the required BBN energies include potential model calculations as in Refs. \cite{buchmann, nacre2} and calculations using three-cluster Generator Coordinate Method as in Ref. \cite{pierre95}. These calculations show that in the low energy region ($E_{c.m.} = 0-3$ MeV), of relevance for BBN ($T_9=0.01-1$), the astrophysical factor is dominated by several resonances in $^{11}$C. Within the temperature range corresponding to the BBN, the reaction rates have contributions from the astrophysical S-factor originated by the decaying tail of a sub-threshold resonance, by two narrow resonances, and by the tail of high energy resonances in $^{11}$C. Regarding the low energy resonance structure of $^{11}$C, there are only two narrow resonances at excitation energies of $E_x = 8.105$ (c.m. energy = 0.884 MeV) and 8.421 MeV (c.m. energy =  1.376 MeV), respectively, for which both the $\alpha$ and $\gamma$ widths are properly known \cite{hardie}. Recently, in Refs. \cite{freer,yamaguchi} results for the scattering of $\alpha$-particles on $^{7}$Be have been reported and an analysis of the resonant structure of $^{11}$C was performed in the excitation energy range around $8.6-13.8$ MeV. But this energy region contributes to the $^7$Be($\alpha,\gamma$)$^{11}$C reaction rate at high temperatures and have little influence on the rates corresponding to BBN temperatures.   

In this paper, we revisit the $^7$Be($\alpha,\gamma$)$^{11}$C reaction with a potential model method \cite{radcap,Akram_6Li} and study its impact on the $^7$Li abundance using the available structure information of $^{11}$C.  Our work highlights other possible scenarios for this reaction rate. 
  
The paper is organized as follows. In section 2 we explain in brief the potential model formalism and its relation to the radiative capture cross section and to the reaction rates for the A($\alpha,\gamma$)B reaction. In section 3, we discuss our results for the impact of the $^7$Be($\alpha,\gamma$)$^{11}$C reaction rate on the BBN $^7$Li production. Our  summary follows in section 4.

\section{Formalism} \label{sec:forma}

In the potential model, the radiative capture cross sections for the process $A + \alpha \rightarrow B + \gamma$ taking place via an electric transition of multipolarity $L$, are given by the relation \cite{radcap,Akram_6Li},
\begin{eqnarray}
 \sigma &=& 2\,\pi\,\frac{(2\,l_{f}+1)(2\,J_{f}+1)}{(2\,J_{A}+1)(2\,J_{\alpha}+1)}\,\frac{(\hbar\,c)^{3}}{\mu_{\alpha\,A}\,c^{2}}\,\frac{k}{E^{2}}\frac{e^{2}}{\hbar\,c}\, \sum\limits_{J_{i}}(2\,J_{i}+1)\, \nonumber\\
&\times&  \sum\limits_{L}\,\big(Z_{eff(L)}\big)^{2}\, \frac{(L+1)(2\,L+1)}{L}  \frac{k_{\gamma}^{2\,L+1}}{\big((2\,L+1)!!\big)^{2}} 
% \nonumber\\ &\times& 
(\left<l_{f}\,0\,\,L\,0\big|l_{i}0\right> )^{2}   
\left\lbrace \begin{array}{ccc} l_{f} &s&J_{f} \\
 J_{i}&L&l_{i}  \end{array} \right \rbrace^{2} \big|R_{l_{f}\,s\,L\,J_{f}\,l_{i} \,J_{i}} (k) \big|^{2},
\label{a1}
\end{eqnarray}
where $l_i$, is the relative angular momentum in the initial channel with total angular momentum $J_i$ and channel spin $s$. $E = k^2/(2\mu_{\alpha\,A})$ is the initial relative kinetic energy between $A$ and $\alpha$, with $k$ being the corresponding momentum and $\mu_{\alpha\,A}$ the reduced mass.
$k_{\gamma} = (E+ \epsilon)/{\hbar}$ is the momentum of the photon emitted during the transition from the initial state $l_i, s, J_i$ to the final state $l_f, s, J_f$, and $\epsilon$ is the binding energy of the final state $B = A + \alpha$. In Eq. (\ref{a1}), $eZ_{eff(L)}$ is the effective charge for the electric transitions of multipolarity $L$, and is given by
\begin{eqnarray}
eZ_{eff(L)} = e \Big[Z_{\alpha}\Big(\frac{m_A}{m_B}\Big)^L + Z_{A} \Big(-\frac{m_{\alpha}}{m_B}\Big)^L \Big],
\label{a2}
\end{eqnarray}
where $m_i$ and $Z_i$ are the masses and charges of respective particles. $R_{l_{f} s L J_{f} l_{i} J_{i}}(k)$ is the radial overlap integral given by
\begin{eqnarray}
R_{l_{f} s L J_{f} l_{i} J_{i}}(k) = \int_{0}^{\infty}{\rm d}r \,r^{L+2}\, I_{l_{f} s J_{f}}(r)\,\psi_{l_{i} s J_{i}}^{(+)}(k,r),
\label{a3}
\end{eqnarray}
where $r$ is distance between $A$ and $\alpha$. $\psi_{l_{i} s J_{i}}^{(+)}$ is the initial scattering wave function and $I_{l_{f} s J_{f}}(r)$ is the final state radial overlap function. 
In the asymptotic region, the shape of the radial overlap function is governed by the Whittaker function ($W$), i.e., 
\begin{eqnarray}
I_{l_{f}sJ_{f}}(r) \stackrel{r > R_{0}}{\approx} C_{l_{f}sJ_{f}}\,W_{-\eta_f,\,l_{f}+1/2}(2\,\kappa\,r)/r,
\label{a4}
\end{eqnarray}
where $\eta_f$ and $\kappa$ are the Coulomb parameter and the wave number corresponding to the $A-\alpha$ bound state, respectively. $R_0$ is the channel radius beyond which the nuclear interaction between particles $A$ and $\alpha$ becomes negligible. $C_{l_{f}sJ_{f}}$ is the asymptotic normalization coefficient (ANC), here given in units of fm$^{-1/2}$, for the virtual decay $B \rightarrow A + \alpha$.  

The radial integral $R$ is the most important quantity to obtain the radiative capture cross section,  involving calculations of the radial overlap function $I_{l_{f} s J_{f}}(r)$ and the scattering wave function $\psi_{l_{i} s J_{i}}^{(+)}(k, r)$. To simplify the problem, we will use a two-body potential model to calculate both  $I_{l_{f} s J_{f}}(r)$ and $\psi_{l_{i} s J_{i}}^{(+)}(k, r)$. But notice that the overlap function in Eq. (\ref{a3}) is essentially a many-body object which could  also be  calculated with more elaborated nuclear reaction models. 

In the two-body potential model \cite{radcap,Akram_6Li} the radial overlap function $I_{l_{f} s J_{f}}(r)$ can be expressed in terms of the final bound state wave function  $\phi_{n_{r} l_{f}sJ_{f}}(r)$ as
\begin{eqnarray}
I_{l_{f}sJ_{f}}(r) = S_{n_{r}l_{f}sJ_{f}}^{1/2}\,\phi_{n l_{f}sJ_{f}}(r),
\label{a5}
\end{eqnarray}
where $S_{n_{r}l_{f}sJ_{f}}$ is the spectroscopic factor for the bound state of $B$ in the final channel, $n$ is the principal quantum number and represents the number of nodes in the radial bound state wave function (here we exclude the node at the origin).
For the asymptotic region ($r > R_0$), the bound state wave function in the two body potential model is given by,
\begin{eqnarray}
\phi_{nl_{f}sJ_{f}}(r) \stackrel{r > R_{0}}{\approx} b_{n l_{f}sJ_{f}}\,W_{-\eta_f l_{f}+1/2}(2\,\kappa\,r)/r. 
\label{a6}
\end{eqnarray}
The parameter $b_{n l_{f}sJ_{f}}$ is known as the single-particle ANC and depends on the bound-state potential parameters. 

To calculate the wave functions we use an extended version of the potential model code RADCAP \cite{radcap} and generated the wavefunctions with a Woods-Saxon potential. For some fixed radial ($r_0$) and diffuseness ($a$) parameters, the bound state wave function is obtained by adjusting the potential depth ($V_0$) to reproduce the binding energy. For the continuum wave function the asymptotic behavior is taken in the form
\begin{eqnarray}
\psi_{l_{i}sJ_{i}}^{(+)}(k,r) \approx \frac{e^{-i \delta_{l_{i} s J_{i}}}}{2\,i\,r}\big[I_{l_{i}}(k,r) - e^{2 i \delta_{l_{i}s J_{i}}} O_{l_{i}}(k,r)\big],
\label{a7}
\end{eqnarray}
where $\delta_{l_{i} s J_{i}}$ is the scattering phase shift and $I_{l_{i}}$, $O_{l_{i}}$ are the incoming and outgoing spherical waves, respectively which can be expressed in terms of the regular ($F_{l}$) and singular ($G_{l}$) Coulomb function as, 
\begin{eqnarray}
I_{l_{i}}(k,r)=  G_{l_{i}}(k,r) - i\,F_{l_{i}}(k,r),\nonumber\\
O_{l_{i}}(k,r)= G_{l_{i}}(k,r)  + i\,F_{l_{i}}(k,r).\nonumber
\end{eqnarray} 

For the case of resonances, the potential parameters ($r_0$, $a$, $V_0^c$) are adjusted to reproduce the experimental resonance energies, resonance widths and  scattering phase shifts \cite{Akram_6Li, radcap, shub1}. In this work, we use Eq. (\ref{a5}) to calculate the radial overlap function because the ground state of $^{11}$C is deeply bound and the ANC method can not be used as the capture to ground state is not peripheral.  

From the cross sections [Eq. (\ref{a1})], one can calculate the astrophysical factor using the relation
\begin{eqnarray}
S(E) = E\,e^{2\,\pi\,\eta_{i}}\,\sigma(E),
\label{a8}
\end{eqnarray}
where $\eta_i$ is the Coulomb parameter in the initial channel. The nuclear reaction rate per mole, in general, can be calculated from the cross sections by using the relation
\begin{eqnarray}
N_A\langle \sigma v \rangle &=& N_A\Big(\frac{8}{\pi\,\mu_{\alpha A}}\Big)^{1/2}\, \frac{1}{(k_B\, T)^{3/2}} \int_0^\infty \sigma(E) \, E\, {\rm exp}\Big(-\frac{E}{k_B T}\Big) {\rm d}E,
\label{a9}
\end{eqnarray}
where $k_B$ is the Boltzmann constant and $N_A$ is the Avogadro number. $T$ is the temperature in kelvins and is typically of the order of gigakelvins for astrophysical reactions during the BBN. 

In the case of narrow resonances ($\Gamma << E_r$), the S-factors are calculated using the Breit-Wigner formula,
\begin{eqnarray}
\sigma_r = \frac{\pi}{k^2}\frac{(2J_r+1)}{(2J_A+1)(2J_{\alpha}+1)} \frac{\Gamma_{\alpha}\,\Gamma_\gamma}{(E-E_r)^2+(\Gamma/2)^2}, 
\label{a10}
\end{eqnarray}
where $\Gamma_{\alpha}$ and $\Gamma_\gamma$ are the partial widths in the entrance and decay channel and $\Gamma$ is the total width. $E_r$ and $J_r$ are the resonance energy and the spin of the initial resonance state in nucleus B. Note that in order to calculate the resonance cross section over a wide energy range one has to use energy dependent widths in both channels. These are, e.g., given by Eqs. ($38-40$) of Ref. \cite{shub2}.  The contribution to the reaction rates by very narrow resonances can be approximated as 
\begin{eqnarray}
N_A\langle \sigma v \rangle_r = 1.54 \times 10^{11} \frac{(\omega \gamma)}{(\mu_{\alpha A} T_9)^{3/2}} \exp\left[-{11.605 E_r\over T_9}\right], \nonumber \\
\label{a11}
\end{eqnarray}
where $T_9$ is the temperature in units of $10^9$ K. The quantity $\omega \gamma$ is known as the resonance strength and can be obtained from experiments. It is given by
\begin{eqnarray}
\omega \gamma =\frac{(2J_r+1)}{(2J_A+1)(2J_{\alpha}+1)} \frac{\Gamma_{\alpha}\,\Gamma_{\gamma}}{\Gamma}.
\label{a12} 
\end{eqnarray}
In Eq. (\ref{a11}), we use the reduced mass $\mu_{\alpha A}$ given in atomic mass units and $E_r$ and $\omega \gamma$ given in units of MeV. 

With the results for the reaction rates of the $^7$Be($\alpha, \gamma$)$^{11}$C reaction, we perform BBN calculations using a BBN code based on the Wagoner code \cite{Wag69} and similar to NUC123 \cite{Kaw92} to study the effect of this reaction on the $^7$Li abundance.

\section{Results and discussion} \label{sec:RD}

\subsection{Astrophysical S-factor}
We first discuss the results for the  radiative capture cross section and the astrophysical S-factor for the $^7$Be($\alpha, \gamma$)$^{11}$C reaction in the energy range $0-3$ MeV. In this energy region there are several resonances in $^{11}$C which can contribute to the cross sections. However, for the rate calculations corresponding to BBN temperatures, only the cross sections below 1 MeV are important as the Gamow peak for this reaction lies in the energy range $0.14-0.7$ MeV for temperatures within T$_9$ = $0.1-1$.  
As mentioned earlier, only some theoretical estimates are available for this S-factor \cite{pierre95,buchmann, nacre2} and there is insufficient experimental data.
The latest estimates of the S-factor and rates, which have been used as reference in many studies, are those from the NACRE-II compilation \cite{nacre2}. These are based upon potential model calculations and resonance information (position and total width) extracted from the compilation \cite{kelley}. In the low MeV  energy region ($< 8.5$ MeV) important for the BBN rate calculations only two narrow resonances in $^{11}$C situated respectively at excitation energies of 8.105 MeV (c.m. energy = 0.884 MeV)  and 8.421 MeV  (c.m. energy =  1.376 MeV) are properly known \cite{hardie}, i.e., for which the resonance widths in both channels are known. The S-factor in the low energy region is mainly contributed by the sub-threshold resonance at $E_x$ = 7.50 MeV, the tails of high energy resonances \cite{pierre95,buchmann, nacre2} and by the above mentioned narrow resonances albeit with small contributions.  But, while these two narrow resonances only have a limited effect on the S-factor and on the reaction rate at low temperatures, for the full temperature range of a BBN environment they contribute to the major part of the reaction rate, as shown in Figure \ref{fig_2}.
\begin{table}[ht]
\begin{center}
\caption{Woods-Saxon potential parameters used to calculate the wave function of the respective state. The potential parameters $r_0$, $a_0$ and the spectroscopic factors (SF) for the resonance states are taken from Ref. \cite{nacre2}. \label{ta1}}
\begin{tabular}{|c|c|c|c|c|c|c|}
\hline\hline
$J^{\pi}$ &  {E$_x$ (MeV)} &  {$l_i$}  & {$r_0$ (fm)} & {$a_0$ (fm)} & {$V_0$ or $V_0^c$ (MeV)} & \\
\hline 
${3}/{2}^-$ &  0.0         &  0      &   0.90      &   0.50      & $-68.95$      &  1.0  \\
${5}/{2}^+$ &  8.869       &  1      &   0.70      &   0.40      & $-55.99$      &  0.1  \\
${3}/{2}^-$ &  9.645       &  2      &   0.82      &   0.60      & $-53.96$      &  2.8  \\
${5}/{2}^-$ &  9.780       &  2      &   1.08      &   0.16      & $-34.36$      &  0.23  \\
${7}/{2}^-$ &  9.970       &  2      &   0.93      &   0.10      & $-92.27$      &  1.92  \\
\hline
\hline
\end{tabular}
\end{center}
\end{table}

We would like to remind that in our potential model calculations, apart from the non-resonant contribution, we take into account only those resonances in the above mentioned energy region which decay via electric E1 and E2 transitions. We mainly take into account the 5/2$^+$ ($E_x = 8.8699$ MeV, $E_r = 1.16$ MeV), the 3/2$^-$ ($E_x = 9.645$ MeV, $E_r = 2.11$ MeV), the 5/2$^-$ ($E_x = 9.780$ MeV, $E_r = 2.24$ MeV) and the 7/2$^-$ ($E_x = 9.970$ MeV, $E_r = 2.43$ MeV) resonances in our calculations. The resonance parameters for these states are taken from Ref. \cite{kelley} and we follow Ref. \cite{nacre2} for potential parameters and spectroscopic factors for the different states which have been adjusted to fix the width and position of each resonance. In our calculations we do not take into account the spin-orbit coupling.

\begin{figure}[ht!]
\begin{center}
\includegraphics[width=12cm]{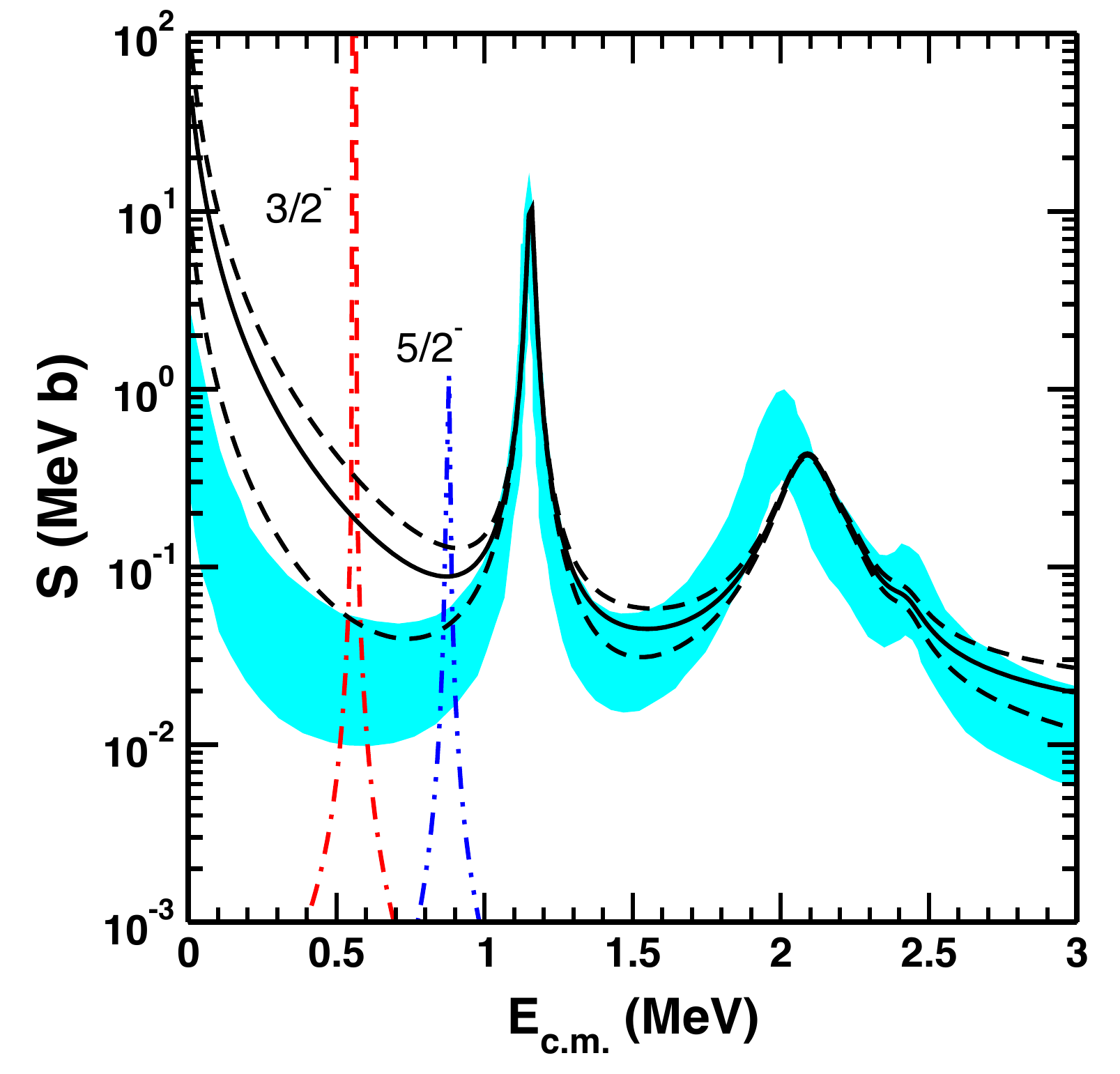}
\caption{Calculated astrophysical S-factor for the $^7$Be($\alpha,\gamma$)$^{11}$C reaction (solid line). The lower and upper dashed lines give the extreme limits of our calculations. The cyan band in the figure represents the limits of S-factors from the NACRE-II compilation \cite{nacre2}. The dot-dashed and double-dot-dashed lines show the S-factor of two narrow resonances in $^{11}$C situated at $E_x = 8.105$ and 8.421 MeV, respectively, calculated using Breit-Wigner parametrizations. For details see text. \label{fig_1}}
\end{center}
\end{figure}

Table I displays the values of Woods-Saxon potential parameters used for the different states. For each resonance, the depth of the potential ($V_0^c$) is adjusted to fix its position which is different in our calculations than in Ref. \cite{nacre2}. But the potential parameters for the non-resonant contribution are kept the same as those in Ref. \cite{nacre2}.
For the sub-threshold resonance 3/2$^+$ ($E_x = 7.5$ MeV, $E_r = -0.045$ MeV) state we take into account the direct capture transition to this state along with the resonant contribution from sub-threshold to ground state. While the direct capture is a transition from $l_i = 0$, the resonant capture is a transition from $l_i = 1$ and therefore these two transitions do not interfere. The direct capture cross sections are calculated using potential depths $V_0^c = -60$ MeV, $V_0 = -171.16$ MeV with radial and diffuseness parameters being the same as those for the ground state.  
The resonant contributions are obtained using Eq. (\ref{a10}), which requires information about the formation and decay widths. In order to obtain these widths we follow the R-matrix formalism given in Ref. \cite{shub2}, which needs spectroscopic information of the sub-threshold and ground state of $^{11}$C. As these parameters are not known, we use the concept of mirror symmetry and try to deduce them from the mirror nucleus $^{11}$B. Following the method discussed in Ref. \cite{shub2}, we calculate the SF for the 3/2$^+$ ($E_x = 7.977$ MeV) state in $^{11}$B, by fitting its observed radiative width ($\Gamma_\gamma = 0.53$ eV \cite{kelley}) for the E1 transition to the ground state. These calculations also need spectroscopic amplitudes for the ground state, which is obtained by fitting the observed $\Gamma_\gamma$ (0.2 eV) for the E1 decay of 5/2$^+$ resonance at $E_x = 9.271$ MeV \cite{kelley, hardie} to the ground state of $^{11}$B.
Since the transitions in $^{11}$B are not peripheral, we extract the SFs instead of ANCs from our calculations, which for the 3/2$^+$ state, with potential parameters $r_0 = 0.9$ fm and $a_0 = 0.5$ fm, comes out to be about 10. If the mirror symmetry holds then this SF should remain the same for the corresponding 3/2$^+$ state of $^{11}$C with the same potential parameters. We then use this SF to calculate the $\Gamma_\alpha$ and $\Gamma_\gamma$ for the 3/2$^+$ sub-threshold resonance in $^{11}$C, which are then used in Eq. (\ref{a10}) to get the resonant contribution to the S-factor. 
The uncertainties in the S-factors are calculated by varying the SF for ground, SF$_{gs}$, and sub-threshold state, SF$_{st}$, keeping the relative transition probability ($\propto {\rm SF}_{gs}\times  {\rm SF}_{st}$) of these states fixed as 10. Therefore, we use this freedom to vary the individual spectroscopic factors with the range 1-10, keeping their product constant.  The value of $\Gamma_\alpha$ of the sub-threshold resonance calculated at $ \sim 4.6$ keV for a channel radius equal to 9.0 fm comes out to be $5.87\times10^{-72}$ eV if the SF$_{st}$ of this state is taken as 1 and it comes out to be $5.87\times10^{-71}$ eV if SF$_{st}$ = 10. On the other hand, $\Gamma_\gamma$ remains same as 20.85 eV for both cases as it depends upon the spectroscopic factors of both the ground and the sub-threshold state. For rest of calculations in this work, we adopted the average values of these parameters, i.e., $\Gamma_\alpha$ as $3.228 \times10^{-71}$ eV and $\Gamma_\gamma$ as 20.85 eV.  

\begin{figure}[ht!]
\begin{center}
\includegraphics[width=12cm]{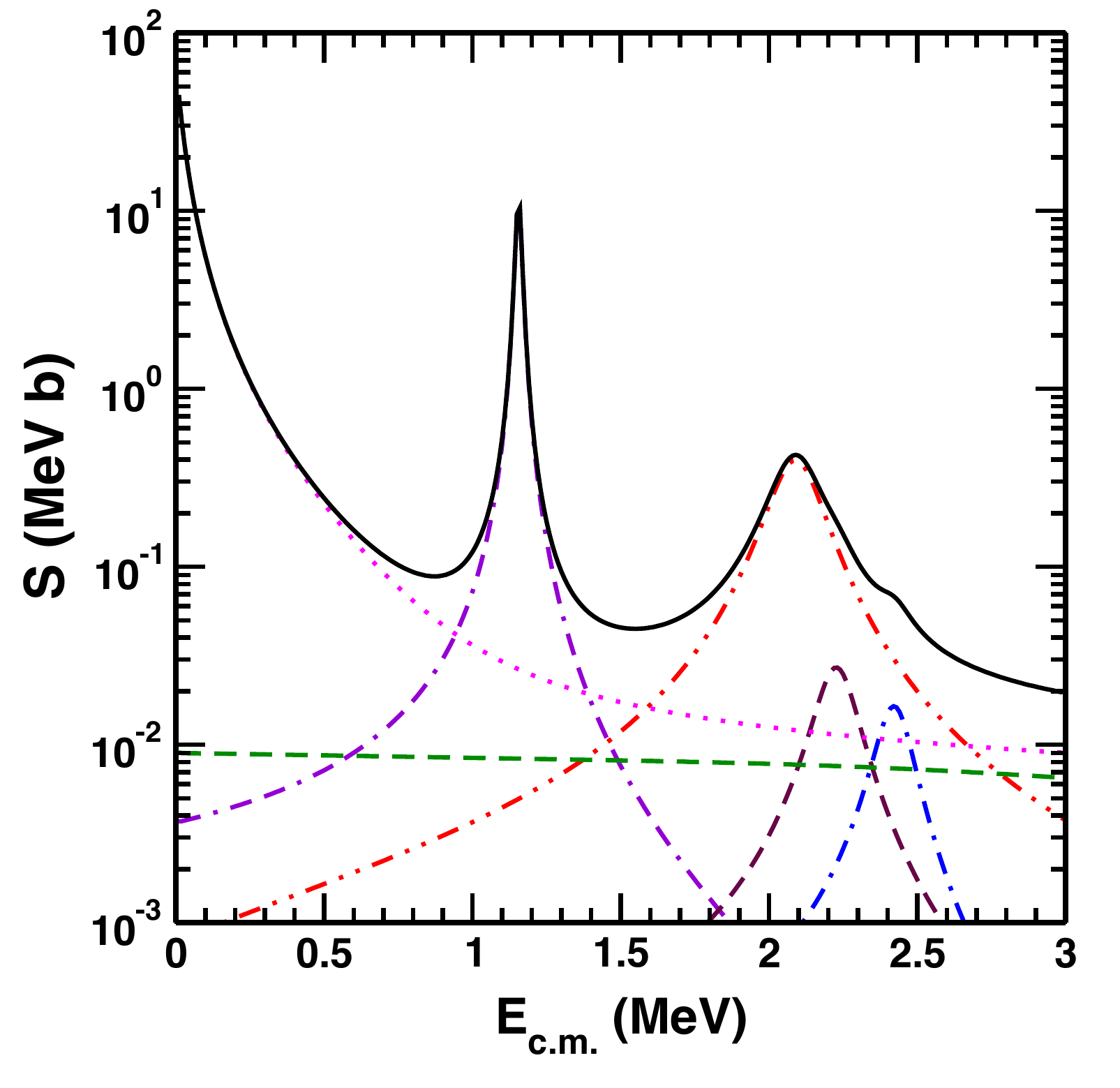}
\caption{Calculated astrophysical S-factor for the $^7$Be($\alpha,\gamma$)$^{11}$C reaction corresponding to the solid line in Fig. \ref{fig_1}. The  S-factor gets  contributions from non-resonant and various resonances displayed separately. The dotted line is due to the central sub-threshold component. The dashed line is for direct capture. The dotted-dashed line (lowest energy resonance) is for the $5/2^+$ resonance at $E_x = 8.8699$ MeV ($E_r = 1.16$ MeV). The dotted-dotted-dashed line is for the $3/2^-$  resonance at $E_x = 9.645$  MeV ($E_r=2.11$ MeV). The dashed curve  is for the $5/2^-$ resonance at $E_x = 9.780$ MeV ($E_r = 2.24$ MeV) and the dotted-dashed line is for the $7/2^-$ resonance at $E_x = 9.970$ MeV ($E_r = 2.43$ MeV).
 \label{fig_1b}}
 \end{center}
\end{figure}

In Fig. \ref{fig_1}, we plot our total S-factor (solid line) which is the sum of non-resonant, resonant and sub-threshold resonance contributions. The lower and upper dashed lines gives the limits of our S-factors and correspond to the case when the SF of the 3/2$^+$ state is taken as either 1 or 10, respectively. We compare our calculations with that from NACRE-II. The cyan band in the figure shows the limits of the S-factor from the NACRE-II compilation \cite{nacre2}, where the contribution of sub-threshold state is taken from Ref. \cite{pierre95}. We also plot the S-factors of the two narrow resonances at $E_r = 0.560$ (dot-dashed line) and $0.877$ MeV (double-dot-dashed line), respectively, calculated with the Breit-Wigner formula. 

In Fig. \ref{fig_1b} we break down the contributions contributions from non-resonant and various resonances for our calculated astrophysical S-factor for the $^7$Be($\alpha,\gamma$)$^{11}$C reaction corresponding to the solid line in Fig. \ref{fig_1}. The  S-factor gets  contributions from non-resonant and various resonances displayed separately. The dotted line is due to the central sub-threshold component. The dashed line is for direct capture. The dotted-dashed line (lowest energy resonance) is for the $5/2^+$ resonance at $E_x = 8.8699$ MeV ($E_r = 1.16$ MeV). The dotted-dotted-dashed line is for the $3/2^-$  resonance at $E_x = 9.645$  Mev ($E_r=2.11$ MeV). The dashed curve  is for the $5/2^-$ resonance at $E_x = 9.780$ MeV ($E_r = 2.24$ MeV) and the dotted-dashed line is for the $7/2^-$ resonance at $E_x = 9.970$ MeV ($E_r = 2.43$ MeV).

\subsection{Reaction rates and the $^7$Li abundance}

\begin{figure}[ht!]
\begin{center}
\includegraphics[width=12cm]{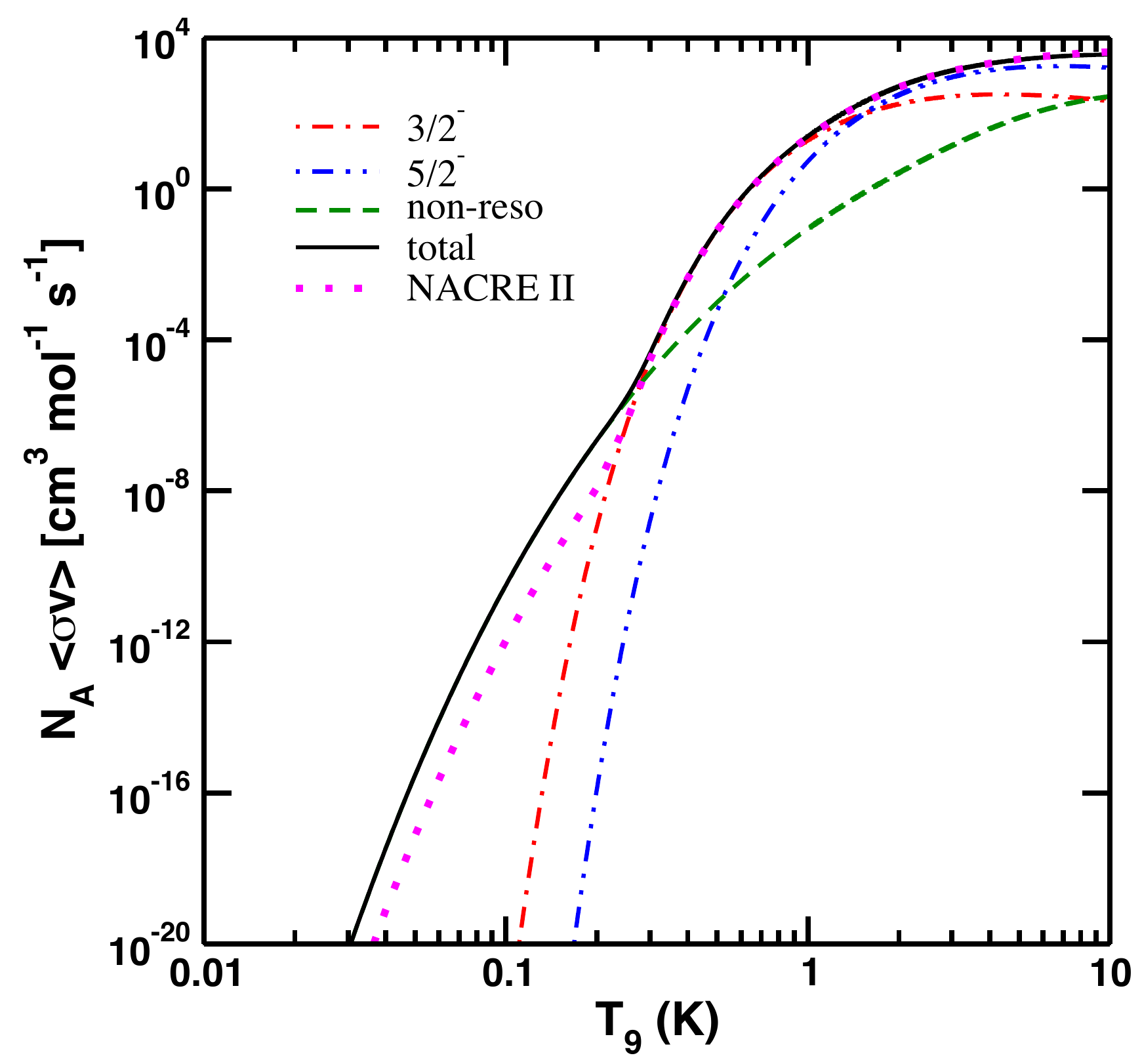}
\caption{{Total reaction rate (solid black line) of $^7$Be($\alpha,\gamma$)$^{11}$C reaction in the temperature range $10^7 - 10^{10}$ K. We also plot individual contributions of two low-lying narrow resonances along with the non-resonant rate (non-reso). The contribution of the $3/2^+$ sub-threshold state is included in our non-resonant rate. The dotted line shows the reaction rate published by NACRE II \cite{nacre2}.}  \label{fig_2}}
 \end{center}
\end{figure}

The S-factor obtained above (solid line in Fig. \ref{fig_1}) is then used as an input in Eq. (\ref{a9}) to calculate the reaction rate. In order to get the total reaction rate we also add the contributions of the two narrow resonances calculated using Eq. (\ref{a11}). In Figure \ref{fig_2}, we plot our total reaction rate along with the non-resonant and resonant contributions of two narrow resonances, $3/2^-$ at $E_r = 0.560$ and 5/2$^-$ at $0.877$ MeV, respectively. The non-resonant rate includes contributions from direct capture and capture through the sub-threshold resonance. It is clear from the figure that the total rate is mainly contributed by the non-resonant rate up to the temperature $2.5\times10^{8}$ K. Beyond this temperature, the $3/2^-$ resonance governs the reaction rate. Around the temperature of $1\times10^{9}$ K the $5/2^-$ resonance at $0.877$ MeV also starts contributing significantly along with the other high energy resonances listed in Table \ref{ta1}. Beyond the temperature $1.5\times10^{9}$ K, the narrow $5/2^-$ resonance starts dominating over the $3/2^-$ resonance and their difference increases significantly with temperature. Figure 2 clearly indicates that the two experimentally observed resonances play a major role  in the calculation of the reaction rate. It is also clear from the figure that for temperatures in the range important for BBN  (up to about $1\times10^9$ K) contributions to the rate from higher resonances are very small.

 After calculating the total rates, we include these in the BBN network calculations to study the effect of $^7$Be($\alpha,\gamma$)$^{11}$C reaction on the $^7$Li abundance. We use $\eta_B =  6.1 \times 10^{-10}$ and the neutron lifetime equal to $880$ sec. Our preliminary finding is that the $^7$Be($\alpha,\gamma$)$^{11}$C reaction calculated using the available information on various low energy resonances, has no appreciable effect on the $^7$Li abundance. Considering the uncertainty of the spectroscopic factor used \cite{nacre2}, we have also tried to increase the reaction rates, specially in the low energy region, to check if any significant changes arise in the $^7$Li abundances due to increase in $^7$Be($\alpha,\gamma$)$^{11}$C reaction rates. It is worth noticing that multiplying these reaction rates with a large number, say $10^4$, does not yield any significant change in the $^7$Li abundance. This ensures that with the present spectroscopic information of $^{11}$C, there is no apparent solution of the lithium problem arising from significant variations of the  $^7$Be($\alpha,\gamma$)$^{11}$C reaction rate.

\section{Summary}
In summary, we have studied the $^7$Be($\alpha,\gamma$)$^{11}$C reaction using a two-body potential model \cite{radcap, Akram_6Li} in order to assess its impact on the primordial $^7$Li abundance. As with other theoretical models, we have shown that it is basically impossible to generate low energy resonances in $^{11}$C of relevance for the big bang nucleosynthesis. For the reaction rate calculations corresponding to BBN temperatures, the low energy region ($E_r = 100-700$ keV) of the S-factor has contributions from the decaying tail of the sub-threshold resonance at $E_r = -0.045$ MeV and by the tails of a few high energy resonances. Given the unavailability of observed spectroscopic information on the sub-threshold state, we extract these from the mirror nucleus $^{11}$B and calculate the S-factor for this state which comes out significantly larger than those reported in Ref. \cite{pierre95} obtained with the resonating group method. 

The reaction rates calculated  for the $^7$Be($\alpha,\gamma$)$^{11}$C reaction has no impact on the primordial $^7$Li abundance. Increasing such rates by an absurdly large factor (namely, about $10^4$ times) does not yield any significant change in the $^7$Li abundance. 

\bigskip

{\bf Acknowledgements}

C.A.B. acknowledges support from the U.S. NSF Grant number 1415656 and the U.S. DOE Grant number DE-FG02-08ER41533. A.M.M. acknowledges support from the U.S. DOE Grant number DE-FG02-93ER40773, NNSA Grant number DE-NA0003841 and the U.S. NSF Grant number PHY-1415656. S. H. acknowledges support from the Grant number 11705244 of NNSF China. We also acknowledge support by the CUSTIPEN (China-U.S. Theory Institute for Physics with Exotic Nuclei) under DOE Grant No. DE-FG02-13ER42025.


\begin{thebibliography}{99}
\bibitem{komatsu} E. Komatsu {\it et al.}, Astrophys. J. Suppl. Ser. {\bf 192}, 18 (2011).
\bibitem{planck}  P. A. R. Ade, N.Aghanim {\it et al.} (Planck Collaboration), Astron. Astrophys. {\bf 571}, A16 (2014).
\bibitem{cyburt} R. H. Cyburt, B. D. Fields and K. A. Olive, Phys. Lett. B {\bf 567}, 227 (2003).
\bibitem{coc04} A. Coc, {\it et al.}, Astrophys. J. {\bf 600}, 544 (2004).
\bibitem{sbordone} L. Sbordone {\it et al.}, Astron. Astrophys. {\bf 522} A26 (2010).
\bibitem{pospelov} M. Pospelov and J. Pradler Annu. Rev. Nucl. Part. Sci {\bf 60}, 539 (2010).
\bibitem{coc13} A. Coc, {\it et al.}, Phys. Rev. D {\bf 87} 123530 (2013).
\bibitem{kusakabe} M. Kusakabe, K. S. Kim, M. K. Cheoun {\it et al.}, Astrophys. J {\bf 214} 5 (2014).
\bibitem{yamazaki} D. G. Yamazaki, M. Kusakabe, {\it et al.}, Phys. Rev. D {\bf90}, 023001 (2014).
\bibitem{goudelis} A. Goudelis, M. Pospelov and J. Pradler, Phys. Rev. Lett. {\bf 116} 211303 (2016).
\bibitem{suqing1} S.Q. Hou, J.J. He, A. Parikh, D. Kahl, C.A. Bertulani, T. Kajino, G.J. Mathews and G. Zhao,  Astrophys. J. {\bf 834}, 165 (2017).
\bibitem{daniel} { D. Cumberbatch {\it et al.}, Phys. Rev. D {\bf 76}, 123005 (2007).}
\bibitem{reeves} H. Reeves, W. A. Fowler, and F. Hoyle, Nature {\bf226}, 727 (1970).
\bibitem{Lam17} L. Lamia, et al., Astrophys. J. 850, 175 (2017)
\bibitem{coc10} A.Coc and E. Vangioni, J. Phys.: Conf. Ser. {\bf 202}, 012001 (2010).
\bibitem{broggini12} C. Broggini {\it et al.}, J. Cosmol. Astropart. Phys. 06, 030 (2012).
\bibitem{boer} R. J. deBoer {\it et al.}, Phs. Rev. C {\bf 90}, 035804 (2014).
\bibitem{coc17} A.Coc and E. Vangioni, Int. J. Mod. Phys. E {\bf 26}, 1741002 (2017).
\bibitem{hou} S.Q. Hou {\it et al}, Phys. Rev. C {\bf 91} 055802 (2015).
\bibitem{barba} M. Barbagallo {\it et al.}, Phys. Rev. Lett. {\bf 117} 152701 (2016).
\bibitem{kawa} T. Kawabata {\it et al.}, Phys. Rev. Lett. {\bf 118} 052701 (2017).
\bibitem{lamia} L. Lamia {\it et al.}, Astrophys. J. {\bf 850} 175 (2017).
\bibitem{civit} O. Civitarese, M. E. Mosquera, Nucl. Phys. A {\bf 898}, 1 (2013).
\bibitem{hammache} F. Hammache {\it et al.}, Phys. Rev. C 88, 062802(R) (2013).
\bibitem{buchmann} L. Buchmann, J. M. D'Auria, P. McCorquodale, Astrophys. J. {\bf324}, 953 (1988).
\bibitem{nacre2} Y. Xu {\it et al.}, Nucl. Phys. A {\bf 918}, 61 (2013).
\bibitem{pierre95} P. Descouvemont, Nucl. Phys. A {\bf584} 532 (1995).
\bibitem{hardie}G. Hardie, B. W. Filippone, A. J. Elwyn, M. Wiescher and R. E. Segel, Phys. Rev. C {\bf 29}, 1199 (1984).
\bibitem{freer} M. Freer {\it et al.}, Phys. Rev. C {\bf 85}, 014304 (2012).
\bibitem{yamaguchi} H. Yamaguchi {\it et al.}, Phys. Rev. C {\bf 87}, 034303 (2013).
\bibitem{radcap} C. A. Bertulani, Comp. Phys. Comm. {\bf 156}, 123 (2003).
\bibitem{Akram_6Li} A. M. Mukhamedzhanov, Shubhchintak and C. A. Bertulani, Phys. Rev. C {\bf 93}, 045805 (2016).
\bibitem{shub1} Shubhchintak, C. A. Bertulani, A. M. Mukhamedzhanov and A. T. Kruppa, J. Phys. G: Nucl. Part. Phys. {\bf 43}, 125203 (2016).
\bibitem{shub2} A. M. Mukhamedzhanov, Shubhchintak, C. A. Bertulani and T. V. Nhan Hao, Phys. Rev. C {\bf 95}, 024616 (2017).
\bibitem{Wag69} R.V. Wagoner, Ap. J. Suppl. Ser. {\bf 18}, 247 (1969).
\bibitem{Kaw92} L. Kawano, ``Let's Go: Early Universe. Primordial Nucleosynthesis: The Computer Way'';
NASA Technical Reports Server (NTRS): Hampton, VA, USA, 1992.
\bibitem{kelley} J. H. Kelley, E. Kwan, J. E. Purcell, C. G. Sheu, H. R. Weller, Nucl. Phys. A {\bf 880}, 88 (2012).


\end{thebibliography}
\end{document}